\documentclass[11pt,twoside,a4paper]{article}
\usepackage{datetime}\usdate
\usepackage{graphicx}
\usepackage{artmods,macros}

\usepackage[latin1]{inputenc}  %swedish letters
\usepackage[T1]{fontenc}      %svenska avstavningsregler o dyl

\usepackage{amsmath,amssymb}
\usepackage{subfigure}
\usepackage{color}
\usepackage{cite}

\usepackage[font=small,labelfont=bf,labelsep=period,margin=0.5cm]{caption}
\usepackage{booktabs} % cmidrule
\usepackage{dcolumn}% Align table columns on decimal point
\usepackage{enumerate}

\usepackage[flushleft]{threeparttable}

\usepackage{bm}% bold math
\usepackage{hyperref}
\definecolor{darkred}{rgb}{0.5,0,0}
\definecolor{darkgreen}{rgb}{0,0.5,0}
\definecolor{darkblue}{rgb}{0,0,0.5}
\definecolor{red}{rgb}{1,0,0}
\definecolor{green}{rgb}{0,1,0}
\definecolor{blue}{rgb}{0,0,1}
\hypersetup{ colorlinks,
linkcolor=blue,
filecolor=green,
urlcolor=red,
citecolor=blue }

\usepackage{lineno}
\input macros

\providecommand{\abs}[1]{\left\lvert #1 \right\rvert}
\providecommand{\norm}[1]{\lVert#1\rVert}

\def\Re{\mathbb{R}}

\def\p{\%}
\def\E{\mathcal{E}}

\def\PTV{\mathcal{P}}
\def\CTV{\mathcal{C}}
\def\dsdca{\tilde d}

\def\A{\mathcal{A}}
\def\B{\mathcal{B}}
\def\F{\mathcal{F}}

\def\OO{\mathcal{O}}

\def\M{\mathcal{M}}
\def\K{\mathcal{K}}
\def\P{\mathcal{P}}
\def\R{\mathcal{R}}
\def\S{\mathcal{S}}

\def\Dv{v}

\def\D50{\textrm{D}_{50}}

\def\dA{d_{\textrm{A}}}
\def\dB{d_{\textrm{B}}}

\def\p{\%}

\author{Albin FREDRIKSSON\thanks{E-mail: {\tt albin.fredriksson@raysearchlabs.com}}~$^\dagger$ and Rasmus BOKRANTZ\thanks{RaySearch Laboratories, Sveav\"{a}gen 44, SE-111 34 Stockholm, Sweden.}}

\title{The scenario-based generalization of radiation therapy margins}
   \date  { %April, 2013
\today
          }

\begin{document}
%%%%%%%%%%%%%%%%%%%%%

\maketitle
\vspace{-0.5cm}
\begin{abstract}
\noindent
We give a scenario-based treatment plan optimization formulation that is equivalent to planning with geometric margins if the scenario doses are calculated using the static dose cloud approximation. If the scenario doses are instead calculated more accurately, then our formulation provides a novel robust planning method that overcomes many of the difficulties associated with previous scenario-based robust planning methods. In particular, our method protects only against uncertainties that can occur in practice, it gives a sharp dose fall-off outside high dose regions, and it avoids underdosage of the target in ``easy'' scenarios. The method shares the benefits of the previous scenario-based robust planning methods over geometric margins for applications where the static dose cloud approximation is inaccurate, such as irradiation with few fields and irradiation with ion beams. These properties are demonstrated on a suite of phantom cases planned for treatment with scanned proton beams subject to systematic setup uncertainty. 

\end{abstract}

\section{Introduction}
Safety margins are the most utilized means of generating radiation therapy treatment plans that are robust against errors. For instance, to mitigate the risk that the target is located differently during the plan delivery than during the image acquisition, a planning target volume (PTV) is usually defined as a geometric expansion of the clinical target volume (CTV), and planning is performed to deliver high dose to the full PTV. Analogously, the uncertainty in the position of an organ at risk (OAR) can be mitigated by an expansion into a planning organ-at-risk volume (PRV). Recipes for the calculation of margins on the basis of systematic and random uncertainties and the amplitude of respiration are surveyed in van Herk~\cite{vanherk04}.

An underlying assumption when geometric margins are used is that the static dose cloud approximation is accurate. According to this approximation, the patient is moving (as an effect of the errors) inside a static dose distribution that is not affected by the motion~\cite{unkelbach09}. This approximation is often sufficiently accurate for tumors in regions of relatively homogeneous density, such as cancers in the pelvic region~\cite{craig03}. The approximation is, however, inaccurate for sites of heterogeneous density, such as lung, which can lead to under- or overdosage if the colocation of the target and the treatment fields differs from what was planned~\cite{craig03}. The approximation is also inaccurate for treatments with few fields, because it does not model that a setup shift in parallel with a beam axis only affects the corresponding beam dose marginally (mainly by scaling according to the inverse-square law), whereas perpendicular shifts affect the beam dose to a large extent (because the dose is displaced). The situation is aggravated for ion treatments, the depth-dose curves of which are more sensitive to the density distribution of the treated volume than those of photon treatments~\cite{lomax08b}.

As alternatives to geometric margins, methods that take uncertainties into account explicitly in the optimization have been proposed~\cite{unkelbach07, pflugfelder08, unkelbach09, fredriksson11}. In these methods, dose distributions for multiple error scenarios (e.g., setup shifts, scaled densities, or displaced organs) are computed and the treatment plans are optimized with respect to all of these doses simultaneously. The methods differ in how they take the dose distributions into account: expected value optimization~\cite{unkelbach07, unkelbach09} minimizes the expectation of the objective value over the scenario doses, composite worst case optimization~\cite{fredriksson11} minimizes the objective value of the worst case scenario, and voxelwise worst case optimization~\cite{unkelbach07, pflugfelder08} minimizes the objective function applied to a worst case dose distribution defined as the worst scenario dose to each voxel considered independently. While these methods can produce robust plans also for sites of heterogeneous density, they result in qualitatively different plans than margin-based planning, even in situations where the static dose cloud approximation holds. The differences are not always desirable~\cite{fredriksson12, fredriksson14}: expected value optimization protecting against systematic errors results in gradual dose fall-offs, something that only marginally increases the probability of tumor control but still results in highly increased integral dose; composite worst case optimization risks neglecting ``easy'' scenarios; and voxelwise worst case optimization does not utilize that an increased dose and the periphery of the high-dose region leads to a steeper dose gradient against the non-involved tissue and is overly conservative when used in combination with dose-volume histogram (DVH) goals, because it considers an unphysical dose distribution.

In the present paper, we introduce a generalization of margin-based planning. The generalization is scenario-based, but does not suffer from the disadvantages mentioned above. When the static dose cloud approximation is exact, our method is equivalent to planning with geometric margins. However, the method can also be used to achieve robust plans when the static dose cloud approximation is inaccurate. Because the static dose cloud approximation is implicitly assumed with the use of geometric margins, this implies that our method is a generalization of treatment plan optimization with margins.

\section{Methods}
We generalize the concept of treatment plan optimization with margins to the case when scenario doses are computed for multiple scenarios. The following exposition concerns a single CTV subject to setup errors, but the method can be readily applied to multiple regions of interest (ROIs), including OARs. For simplicity of the presentation, it is assumed that all setup error scenarios correspond to geometric shifts that can be represented by vectors with all components being multiples of the voxel dimensions, and that the voxels of the considered ROI lines up with the dose grid.

The exposition is structured as follows: We first highlight that the use of margins entails the implicit assumption that the static dose cloud approximation holds. We then show how a voxel-separable plan evaluation criterion applied to a PTV can be reformulated as a criterion applied to the CTV under multiple scenarios with scenario doses given by the static dose cloud approximation. Using this reformulation, we define a scenario-based criterion that utilizes the true, non-approximate, scenario doses---this is the scenario-based generalization of margins. Finally, we give a number of possible refinements of this criterion and discuss how to adapt the method to other criteria than directly voxel-separable ones.

\subsection{Margins implicitly assume the static dose cloud approximation}
Target margins are generally defined using a geometric expansion of the CTV, and planning is performed to deliver high dose to the PTV, which is the union of the CTV and the margin~\cite[Section 4.5]{icru83}. The size of the PTV is selected so that the probability that the CTV will be within the PTV is high~\cite{vanherk00}. For this to imply a high probability of the CTV receiving a high dose---which is the ultimate goal of the use of a margin---the static dose cloud approximation must be reasonably accurate: If it is, then whenever the CTV displaced by the errors is geometrically within the PTV, it will receive a high dose. If the approximation on the other hand is inaccurate, meaning that the dose might deform as an effect of the displacement, then there are no guarantees regarding the dose received by the moving CTV. This is the reason behind the failure of geometric margins in highly modulated proton therapy~\cite{lomax08b, fredriksson11}.

\subsection{A scenario-based reformulation of margins}\label{sec-reformulation}
We consider an arbitrary plan evaluation criterion that is additively separable over the voxels. Let the voxelwise component of the criterion be given by the function $\phi$. Applied to the PTV, the criterion takes the form
\begin{equation}\label{ptv}
\sum_{i \in \PTV} \phi(d_i),
\end{equation}
where $d_i$ is the dose to voxel $i$ and $\PTV$ an index set over the PTV voxels. Common voxel-separable criteria include minimum, maximum, and uniform dose penalties, see, e.g., Oelfke and Bortfeld~\cite{oelfke01}.

\begin{figure}[hbtp]
\centering
\subfigure[A patient with a CTV and a PTV.]{
\includegraphics[scale=0.82]{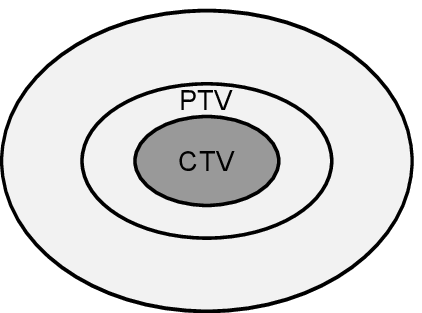}
}
\quad
\subfigure[The dashed ellipses illustrate $\CTV(s) = \{ i + j(s) : i \in \CTV \}$ for two scenarios $s$.]{
\includegraphics[scale=0.82]{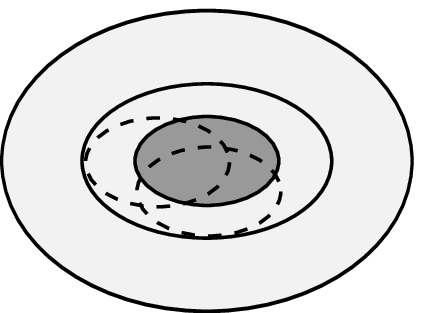}
}
\quad
\subfigure[The set $\cup_{s \in \S} \CTV(s)$ approaches $\P$ as the number of scenarios increases.]{
\includegraphics[scale=0.82]{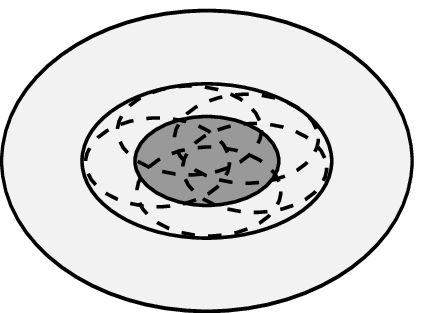}
}
\caption{Construction of the PTV from translated copies of the CTV.}
\label{fig-ptv2}
\end{figure}

\begin{figure}[hbtp]
\centering
\includegraphics[scale=0.8]{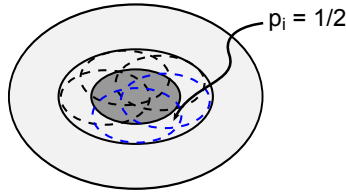}
\caption{Voxel $i$ has weight \mbox{$p_i = 1/2$} because it is contained in $\CTV(s)$ (dashed ellipses) under two scenarios $s$.}
\label{fig-overlap}
\end{figure}

According to the static dose cloud approximation, the effect of a setup error is that the patient is shifted rigidly inside a static dose distribution~\cite{unkelbach09}. A rigid shift of the patient corresponds to the voxel indices being shifted by a fixed offset (assuming that the shift lines up with the dose grid). Let $\S$ denote the set of scenarios and let the offset corresponding to the shift $s$, for $s$ in $\S$, be denoted by $j(s)$. The CTV voxels under scenario $s$ are then given by the scenario-dependent set \mbox{$\CTV(s) := \{ i + j(s) : i \in \CTV \}$}, where $\CTV$ enumerates the CTV voxels in the nominal scenario (corresponding to no error). Because the PTV is an expansion of the CTV, it holds that \mbox{$\PTV = \cup_{s \in \S} \CTV(s)$} if $\S$ is appropriately defined, as illustrated in Figure~\ref{fig-ptv2}.
The criterion~\eqref{ptv} can thus be equivalently stated as
\[
\sum_{i \in  \bigcup_{s \in \S} \CTV(s)} \phi(d_i).
\]
This expression, in turn, is equivalent to
\[
\sum_{s \in \S} \sum_{i \in \CTV(s)} p_{i} \phi(d_i),
\]
where $p_{i}$ is the reciprocal of the number of scenarios $s$ under which voxel $i$ is in $\CTV(s)$:
\[
p_i = 1 /\left| \left\{ s \in \S : i \in \CTV(s) \right\} \right|,
\]
where $\abs{\cdot}$ denotes the cardinality of a set, see Figure~\ref{fig-overlap}.

Offsetting the voxel indices within a fixed dose distribution is equivalent to keeping the voxel indices fixed but shifting the dose rigidly. The criterion can therefore also be formulated as
\[
\sum_{s \in \S} \sum_{i \in \CTV} p_{i+j(s)} \phi(d_{i+j(s)}).
\]
Let \mbox{$\dsdca_i(s) := d_{i+j(s)}$} denote the dose to voxel $i$ under scenario $s$ and \mbox{$p_{i,s} := p_{i+j(s)}$} denote the corresponding weight. The criterion can then be written as
\begin{equation}\label{ctv}
\sum_{s \in \S} \sum_{i \in \CTV} p_{i,s} \phi(\dsdca_i(s)),
\end{equation}
which is our scenario-based reformulation of the PTV criterion.

In summary, we have derived the following equivalence:
\begin{equation}\label{ptv-equivalence}
\sum_{i \in \PTV} \phi(d_i) = \sum_{s \in \S} \sum_{i \in \CTV} p_{i,s} \phi(\dsdca_i(s)),
\end{equation}
%where $p_{i,s} = 1 / \left| \left\{ s' \in \S : \exists i' \in \CTV : i + j(s) = i' + j(s') \right\}\right|$.
where $\dsdca_i(s) = d_{i+j(s)}$ and
\begin{equation}\label{p-def}
p_{i,s} = 1 / \left| \left\{ s' \in \S : i + j(s) - j(s') \in \CTV \right\}\right|,
\end{equation}
and $\S$ has been selected such that \mbox{$\PTV = \{ i + j(s): i \in \CTV, s \in \S \}$}. %\mbox{$\PTV = \cup_{s \in \S} \CTV(s)$}.

\subsection{The scenario-based generalization of margins}
The expression~\eqref{ctv} provides a reformulation of the criterion~\eqref{ptv} for the PTV to a scenario-based criterion for the CTV. The static dose cloud approximation was used to define the weights $p_{i,s}$ and the dose distribution $\dsdca(s)$. We propose the use the scenario-based criterion~\eqref{ctv} together with the exact dose distribution (or a more accurate approximation than the static dose cloud approximation) of scenario $s$, denoted by $d(s)$, in place of the dose distribution $\dsdca(s)$ estimated by the static dose cloud approximation. In cases where the static dose cloud approximation holds, this means that our formulation will be equivalent to using a PTV; in cases where the static dose cloud approximation does not hold, our formulation provides a novel scenario-based robust planning method.

\subsection{Beam-specific effects}\label{sec-beam-specific}
\begin{figure}[hbtp]
\centering
\subfigure[$\CTV(s)$ under the scenarios $s$ in $\S$.]{
\includegraphics[scale=0.75]{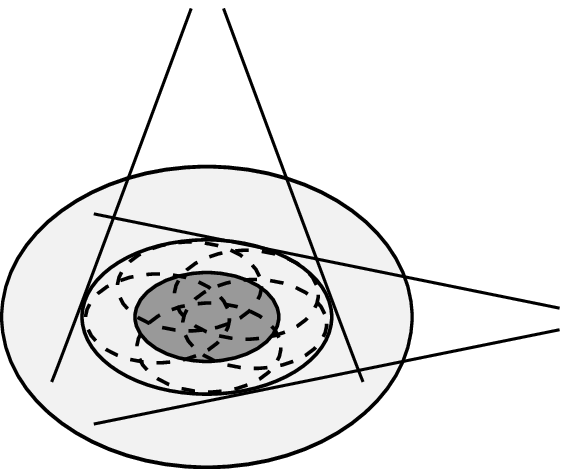}
}
\quad
\subfigure[$\CTV(s)$ projected onto the beam plane for the beam along the x-axis.]{
\includegraphics[scale=0.75]{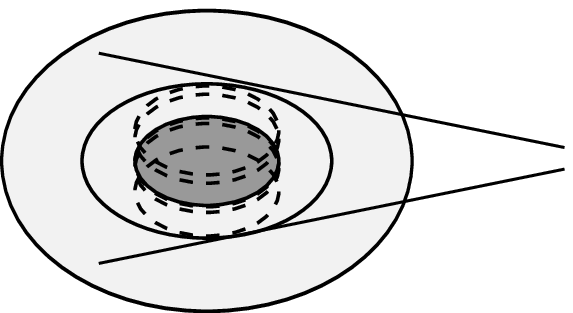}
}
\quad
\subfigure[$\CTV(s)$ projected onto the beam plane for the beam along the y-axis.]{
\includegraphics[scale=0.75]{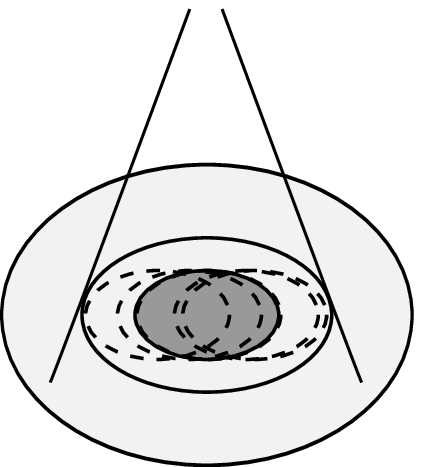}
}
\caption{Projection of $\CTV(s)$ (dashed ellipses) onto the beam planes. The projected sets are used in the calculation of beam-specific weights, but $\phi$ is still evaluated using the scenario doses without projections.}
\label{fig-beam}
\end{figure}

The static dose cloud approximation does not model that the effect of an error can vary with the beams. Such variations can, however, be incorporated in the calculation of the weights $p_{i,s}$ by the definition of weights $p_{i,s}^{(b)}$ for each beam $b$ in the set $\B$ of all beams, for all voxels $i$ and scenarios $s$. The weights $p_{i,s}^{(b)}$ are under this refinement defined according to~\eqref{p-def}, but with $j^{(b)}$ substituted for $j$, where $j^{(b)}(s)$ is the offset in the dose grid corresponding to the shift of scenario $s$ projected onto the beam plane of beam $b$. Thus, $j^{(b)}(s)$ neglects setup errors that are parallel to the beam axis, as illustrated in Figure~\ref{fig-beam}. The weights for different beams are combined into a single weight $p_{i,s}$ by interpolation that takes into account the dose contribution to voxel $i$ under scenario $s$ from each beam. Let $d_i^{(b)}(s)$ denote the dose of beam $b$ delivered to voxel $i$ under scenario $s$. Then, the beam-specific counterpart of the criterion~\eqref{ctv} takes the form
\begin{equation}\label{beam-weighting}
\sum_{s \in \S} \sum_{i \in \CTV} \phi( d_i(s) ) \sum_{b \in \B} \frac{d_i^{(b)}(s)}{d_i(s)} p_{i,s}^{(b)},
\end{equation}
where $d_i^{(b)}(s)/d_i(s)$ is taken as $1/\abs{\B}$ whenever \mbox{$d_i(s)=0$}.

\subsection{Extensions to other types of criteria}
The reformulation (and consequently generalization, when $d$ is substituted for $\dsdca$) of margins can be extended to other types of criteria than additively voxel-separable functions, as the following examples show: For a criterion that is a function $\psi$ of an additively voxel-separable function, such as equivalent uniform dose measures~\cite{niemierko99}, application of $\psi$ to the left- and right-hand side of~\eqref{ptv-equivalence} yields the following reformulation:
\begin{equation*}\label{ptv-equivalence-eud}
\psi \left( \sum_{i \in \PTV} \phi(d_i) \right) = \psi \left( \sum_{s \in \S} \sum_{i \in \CTV} p_{i,s} \phi(\dsdca_i(s)) \right).
\end{equation*}
Another common form of criteria are multiplicatively voxel-separable functions. Criteria of this type include tumor control probabilities, because the overall probability of tumor control is the product of the probability for tumor control in each tumor voxel. Derivations analogous to those in Section~\ref{sec-reformulation} yield the following reformulation for multiplicatively voxel-separable functions:
\begin{equation}\label{ptv-equivalence-tcp}
\prod_{i \in \PTV} \phi(d_i) = \prod_{s \in \S} \prod_{i \in \CTV} \phi(\dsdca_i(s))^{p_{i,s}}.
\end{equation}
For a criterion of the form of a function $\psi$ of a multiplicatively voxel-separable function, such as a normal tissue complication probability calculated using the relative seriality model~\cite{kallman92}, a reformulation is possible by application of $\psi$ to the left- and right-hand side of~\eqref{ptv-equivalence-tcp}. Finally, a constraint for all voxels $i$ in $\PTV$ is equivalent to a constraint for all voxels $i$ in $\CTV(s)$ and scenarios $s$ in $\S$. %For a summary of common treatment plan evaluation criteria, see Romejin et al.~\cite{romeijn04}.

\subsection{Experimental study}
\subsubsection{Optimization formulations and phantom geometries}
We assessed the scenario-based generalization of margins by comparison against geometric margins and against previously published scenario-based robust planning methods. The scenario-based methods that we used for benchmark were (composite) worst case optimization and expected value optimization~\cite[Sections II.D.1--2]{fredriksson12}. The comparison was performed with respect to stylized phantom geometries in either one dimension (1D) or two dimensions (2D), which contained a CTV with voxels enumerated by the set $\CTV$ and, for a subset of the cases, an OAR with voxels enumerated by the set $\OO$. The geometries were planned for treatment with either a single orthogonal proton beam (incident from a dimension orthogonal to the patient geometry) or two coplanar proton beams at a perpendicular angle to each other. Treatment plan optimization using the various methods to achieve robustness was performed with respect to an objective function $f$ penalizing deviation from a uniform dose of one to the CTV and a dose of zero elsewhere, according to
\[
f(d) = w_{\CTV} \sum_{i \in \CTV} \Dv_{i,\CTV} (d_i - 1)^2 + w_{\OO} \sum_{i \in \OO} \Dv_{i,\OO} d^2_i + w_{\E} \sum_{i \in \E} \Dv_{i,\E} d^2_i.
\]
Here, $\E$ enumerates all voxels inside the external contour while for a general structure $\R$, $\Dv_{i,\R}$ is the fraction of the structure that is contained in voxel $i$ (this weight is used to take into account that the ROIs do not necessarily line up with the dose grid), and $w_{\R}$ is a scalar importance weight. Isotropic setup uncertainty of $1$ cm was taken into account in the optimization. A total of four combinations of geometry, beam arrangement, and objective function were considered, which we labeled Examples 1--4, see Table~\ref{tab-examples}.

The examples were selected to clearly exhibit the differences between the methods, and the methods are not necessarily intended to be used for such formulations: For example, the worst case optimization method is not intended to be applied to hopelessly conflicting scenarios, as in Examples 3 and 4. In such cases, a different set of scenarios should ideally be selected for the method to yield the desired result. Moreover, as explained for worst case optimization methods in a previous study~\cite{fredriksson14}, the methods are identical when there are no conflicts.

\begin{table}[htbp]
\caption{Beam arrangement, structure definition, and objective function weights for the example cases. All intervals are specified in centimeters. }
\centering
\label{tab-examples}
\begin{tabular}{l l rrr rrr}
  \toprule
  Example & Beam(s) & $\CTV$ & $\OO$ & $\E$ & $w_{\CTV}$ & $w_{\OO}$ & $w_{\E}$ \\  
  \midrule
  1 & Orthogonal & $[-2,2]$ & -- & $[-6,6]$ & $10$ & --  & $1$ \\
  2 & Coplanar & $\left[ -\frac{3}{2}, \frac{3}{2} \right]^2$ & -- & $[-6,6]^2$ & $100$ &  -- & $1$ \\
  3 & Orthogonal & $\left[ -\frac{3}{2}, \frac{3}{2} \right]^2$ & $[-5,-3] \times [-2,2]$ & $[-6,6]^2$ & $100$ & $100$ & $1$ \\
  4 & Orthogonal & $\left[ -\frac{3}{2}, \frac{3}{2} \right]^2$ & $[-4,-2] \times [-2,2]$ & $[-6,6]^2$ & $100$ & $20$ & $1$ \\
  \bottomrule
\end{tabular}
\end{table}

%% \begin{table}[htbp]
%% \caption{Beam arrangement [coplanar ($\parallelsum$) or orthogonal ($\perp$)], structure definition, and objective function weights for the example cases. All intervals are specified in centimeters. }
%% \centering
%% \label{tab-examples}
%% \begin{tabular}{l r rrr rrr}
%%   \toprule
%%   Ex. & Beams & $\CTV$ & $\OO$ & $\E$ & $w_{\CTV}$ & $w_{\OO}$ & $w_{\E}$ \\  
%%   \midrule
%%   1 & $\perp$ & $[-2,2]$ & -- & $[-6,6]$ & $10$ & --  & $1$ \\
%%   2 & $\parallelsum$ & $\left[ -\frac{3}{2}, \frac{3}{2} \right]^2$ & -- & $[-6,6]^2$ & $100$ &  -- & $1$ \\
%%   3 & $\perp$ & $\left[ -\frac{3}{2}, \frac{3}{2} \right]^2$ & $[-5,-3] \times [-2,2]$ & $[-6,6]^2$ & $100$ & $100$ & $1$ \\
%%   4 & $\perp$ & $\left[ -\frac{3}{2}, \frac{3}{2} \right]^2$ & $[-4,-2] \times [-2,2]$ & $[-6,6]^2$ & $100$ & $20$ & $1$ \\
%%   \bottomrule
%% \end{tabular}
%% \end{table}

\subsubsection{Numerical optimization and dose calculation}
Proton beams were represented by scanning spots selected such that a Bragg peak position matched the voxel center for each voxel in the dose grid within $2$ cm of the CTV. Spot doses for orthogonal beams were calculated using a Gaussian point-spread function with a standard deviation of $3$ mm. Spot doses for coplanar beams were calculated using the analytic proton depth-dose curve in Bortfeld~\cite{bortfeld97}. The dose calculations were performed in a grid with a voxel side length of $1$ mm. The set of possible setup shifts $\S$ was represented by 19 uniformly distributed scenarios in 1D (Example 1) and by nine scenarios (the nominal scenario, all shifts of $1$ cm in the positive and negative axis directions, and all pairwise combinations of the axis directions) in 2D (Examples 2--4). A shift was modeled as a rigid translation of the beam isocenters relative to the patient coordinate system. The spot weights were optimized subject to nonnegativity bounds using the nonlinear programming solver SNOPT v7.2 (Stanford Business Software, Palo Alto, California).% with uniform spot profiles scaled so that the average dose in the CTV equaled one used as a starting point for the optimization.

Treatment plan optimization using scenario-based margins was implemented on the beam-specific form of Section~\ref{sec-beam-specific} for the beam arrangement with two fields (Example 2). Optimization using geometric margins was implemented as minimization of the objective function value in the nominal scenario evaluated with respect geometrically expanded structures. The CTV was expanded to a PTV and the OAR expanded to a PRV by substitution of an expanded volume defined as \mbox{$\{i + j(s): i \in \R, s \in \S\}$} for the original structure $\R$. The objective function weights were scaled with a factor equal to the fractional increase in volume due to the expansion in order to obtain comparable results with the scenario-based methods. The maximum operator used to compute the worst case objective function value in the worst case optimization was approximated by a smooth power mean according to
\[
\max_{s \in S} f(d(s)) \approx \left( \frac{1}{|\S|} \sum_{s \in \S} f(d(s))^{10} \right)^{1/10}
\]
in order for the operator to be continuously differentiable. The expected objective function value used in the expected value optimization was calculated with respect to normally distributed errors with standard deviation selected such that $95\p$ of all values were contained within $1$ cm, conditioned on that an error of at most $1$ cm would occur.

\section{Results}
Doses are quantified in percent of the prescription. We use the term ``margins'' to refer both geometric and scenario-based margins when their results are similar. 

\subsection{Example 1}
This example compares the previous scenario-based methods against margins under very simple conditions: the static dose cloud approximation is exact and there is no OAR that competes with target coverage. The results for geometric and scenario-based margins coincided precisely.

\begin{figure}[hbtp]
\centering
\includegraphics[width=6.5cm]{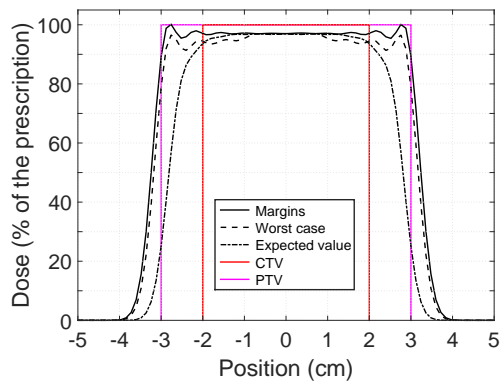}
\caption{Nominal doses for Example 1. The height of the CTV and PTV indicates the prescription dose. }
\label{fig-1d}
\end{figure}

Figure~\ref{fig-1d} shows that all methods gave a uniform dose of about $97\p$ to the innermost $2$ cm of the CTV. Margins kept the planned dose uniform at this level inside the entire PTV, except for at the very edges where the dose was increased to form ``shoulders.'' Worst case optimization created a mildly decreasing dose with increasing distance to the origin outside the innermost $2$ cm of the CTV. Like margins, it created shoulders at the edges of the high-dose region. Expected value minimization created a smooth and gradual fall-off to a dose of about $30\p$ at the edges of the PTV, without any shoulder formations.

\subsection{Example 2}
This example highlights the differences between the scenario-based methods and geometric margins if the static dose cloud approximation is inaccurate. The reason that the static dose cloud approximation gives a poor prediction of a perturbed dose is that the beam doses are unaffected by components of the shifts that are parallel to the beam axis (neglecting the small effect that the beam doses are scaled according to the inverse-square law).

\begin{figure}[hbtp]
\centering
\subfigure[Geometric margins]{
\includegraphics[width=4.3cm]{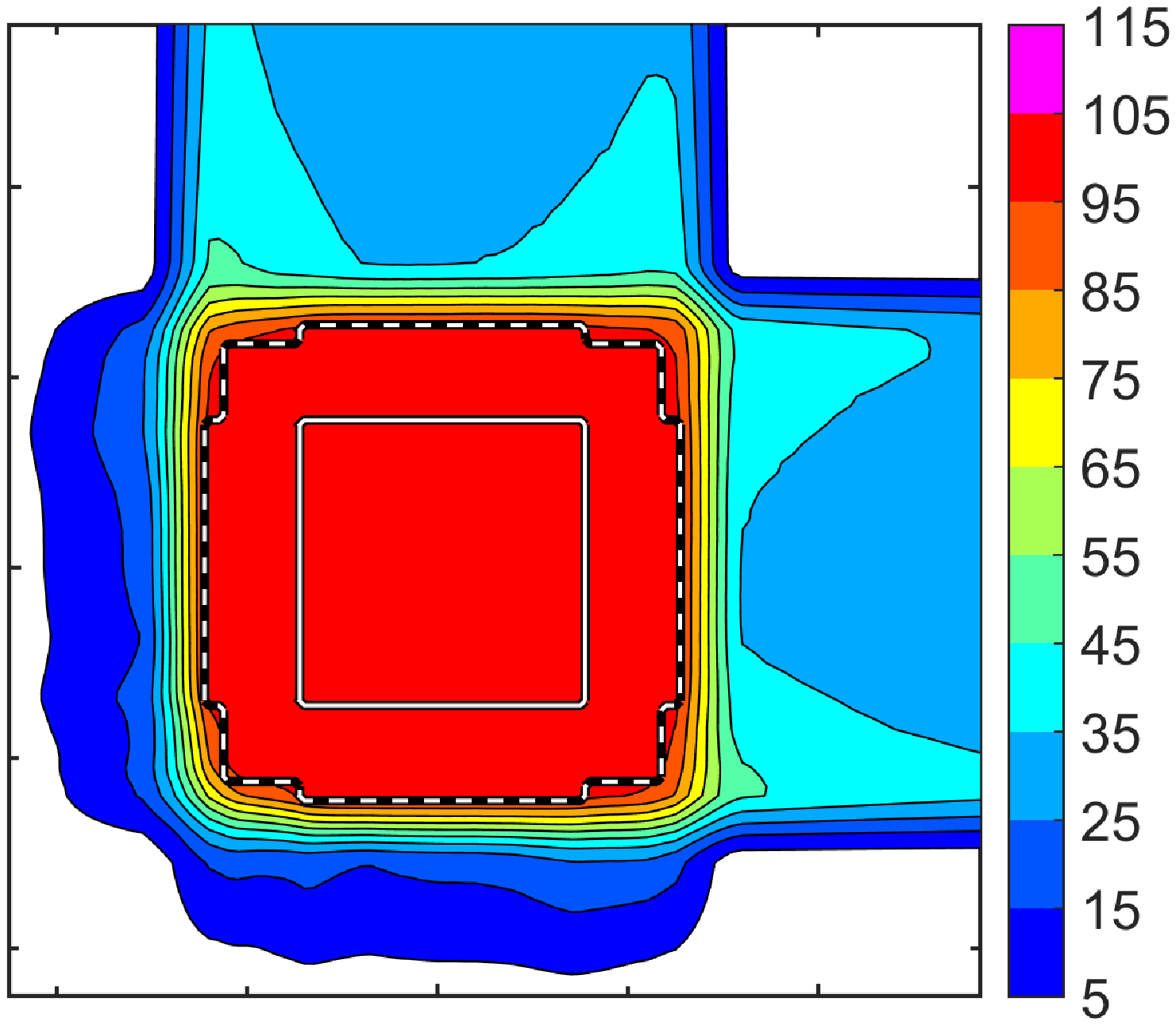}
}
\subfigure[Scenario-based margins]{
\includegraphics[width=4.3cm]{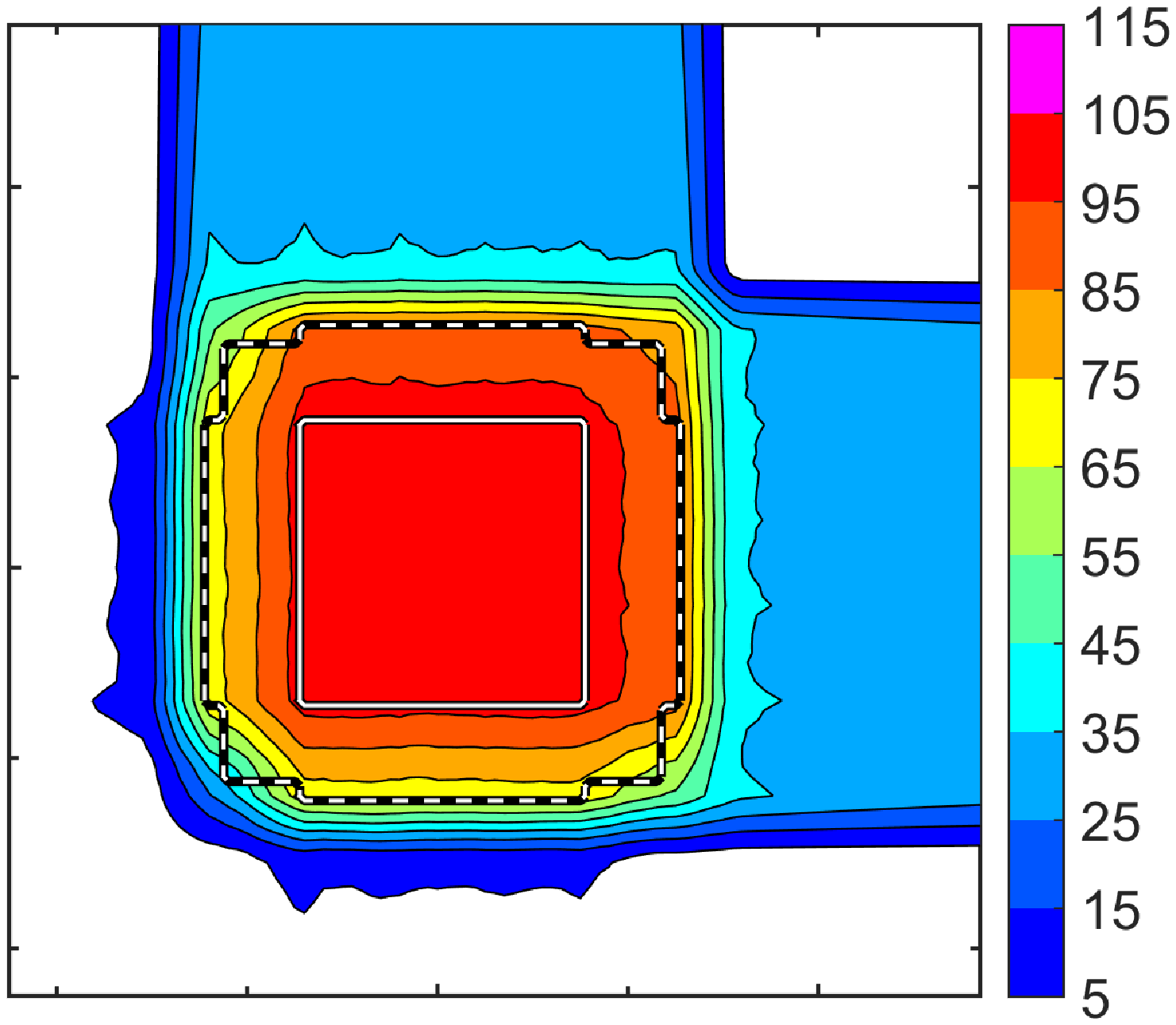}
}
\caption{Nominal doses for Example 2. The CTV is indicated by a solid line and the PTV is indicated by a dashed line.}
\label{fig-2d-1-dose}
\end{figure}

\begin{figure}[hbtp]
\centering
\includegraphics[width=6.5cm]{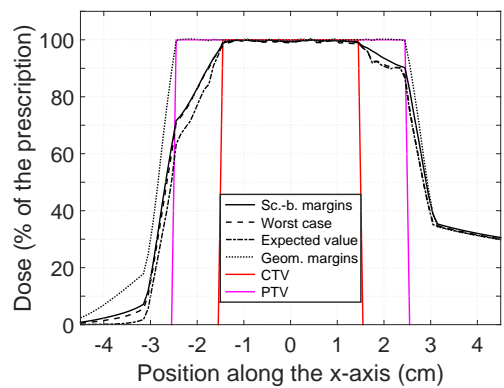}
\caption{Line dose profiles for Example 2 in the nominal scenario for a trace along the x-axis through the isocenter.}
\label{fig-2d-1-linedose}
\end{figure}

\begin{figure}[hbtp]
\centering
\subfigure[Geometric margins]{
\label{fig-2d-1-beamdose-b}
\includegraphics[width=4.3cm]{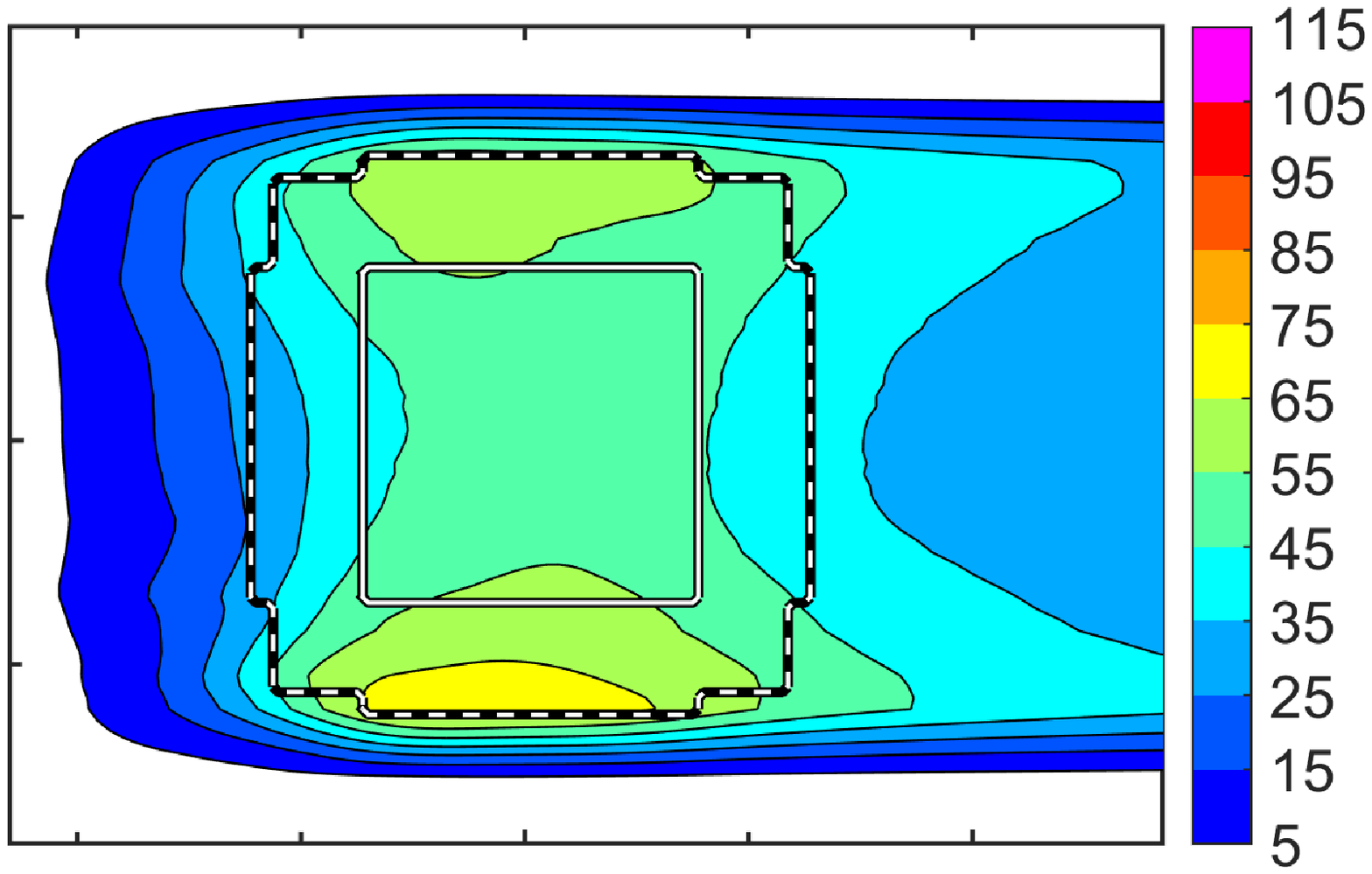}
}
\subfigure[Scenario-based margins]{
\label{fig-2d-1-beamdose-a}
\includegraphics[width=4.3cm]{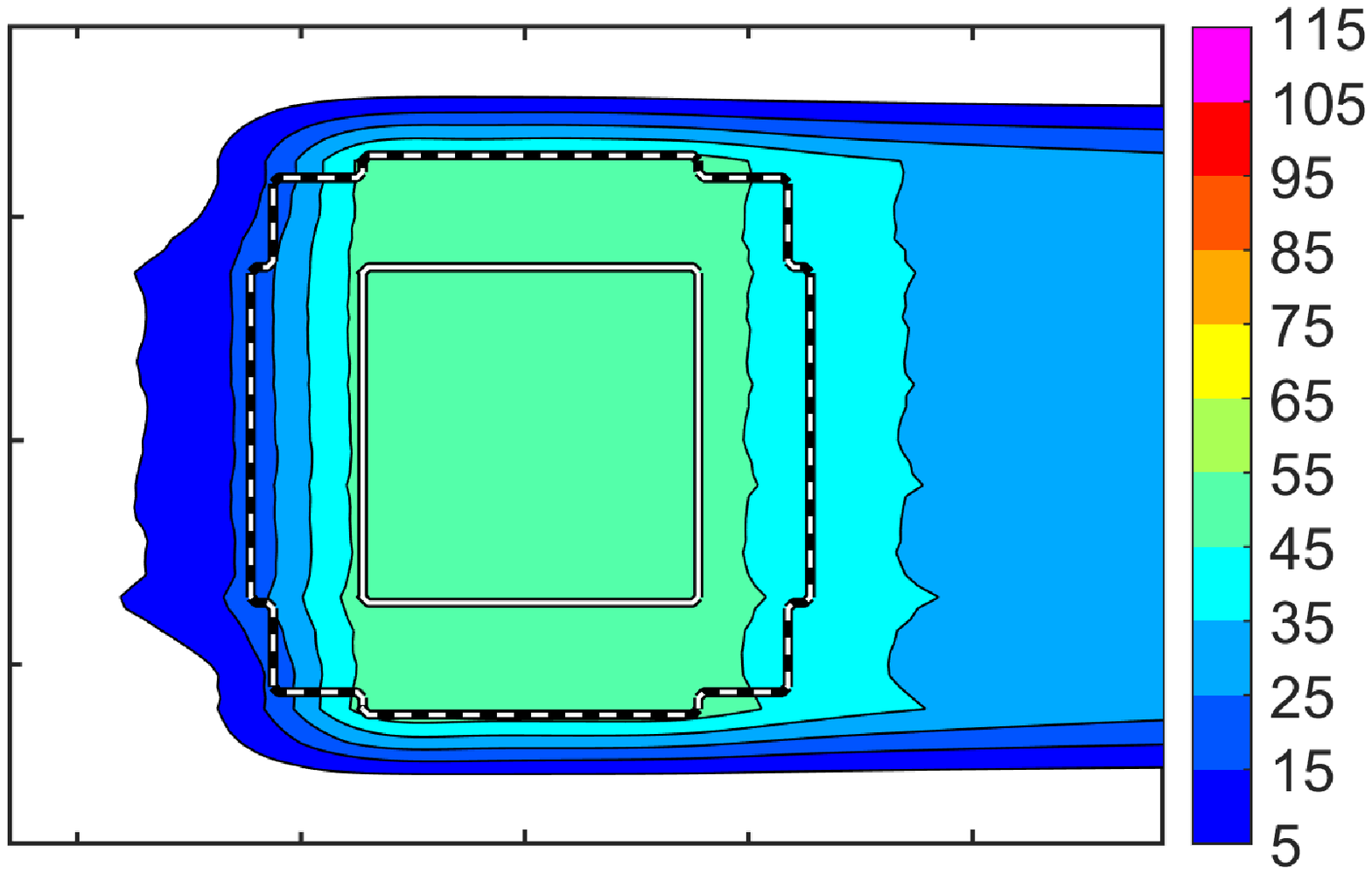}
}
\caption{Beam doses for Example 2 for the beam along the x-axis. The beam doses for the beam along the y-axis were identical except for a $90^\circ$ rotation. The CTV is indicated by a solid line and the PTV is indicated by a dashed line.}
\label{fig-2d-1-beamdose}
\end{figure}

Figure~\ref{fig-2d-1-dose} depicts the nominal dose for scenario-based and geometric margins. Associated DVHs for all scenarios are shown in Figure~\ref{fig-2d-1-dvh}. Corresponding results are not shown for worst case and expected value optimization because of the close similarity with scenario-based margins. However, line doses for all methods are depicted in Figure~\ref{fig-2d-1-linedose}. Geometric margins covered the PTV with a uniform dose at the prescription level. The scenario-based methods instead created a linearly decreasing dose from $100\p$ at the CTV edge to $90\p$ at the proximal edge of the PTV and to about $70\p$ at the distal edge of the PTV. The CTV dose of the scenario-based methods was despite the lower integral dose uniform around $100\p$ in all scenarios, whereas the CTV dose varied in the range $90$--$135\p$ for geometric margins. The beam doses in Figure~\ref{fig-2d-1-beamdose} explain why the plan created by geometric margins is sensitive to perturbations: The saddle-shaped doses (peaks on the lateral edges of the CTV and valleys on the proximal and distal edges) sum to a highly uniform total dose with a steep fall-off in the nominal scenario, but the total dose is irregular if the beams do not patch as intended. The scenario-based methods created uniform beam doses that do not rely on that the beams patch accurately.

\begin{figure}[hbtp]
\centering
\subfigure[Geometric margins]{
\includegraphics[width=4.3cm]{margin_example1_dvh}
}
\subfigure[Scenario-based margins]{
\includegraphics[width=4.3cm]{gm_example1_dvh}
}
%% \subfigure[Minimax]{
%% \includegraphics[width=4.3cm]{mm_example1_dvh}
%% }
%% \subfigure[Expected-value]{
%% \includegraphics[width=4.3cm]{expval_example1_dvh}
%% }
\caption{DVHs for Example 2 in all scenarios.}
\label{fig-2d-1-dvh}
\end{figure}

%% \begin{figure}[hbtp]
%% \centering
%% \subfigure[Conventional margins]{
%% \includegraphics[width=6.5cm]{margin_example1_dose_s0}
%% }
%% %\subfigure[Conventional margins, perturbed dose]{
%% %\includegraphics[width=6.5cm]{margin_example1_dose_s4}
%% %}
%% \subfigure[Conventional margins]{
%% \includegraphics[width=6.5cm]{margin_example1_beamdose_s0}
%% }
%% %% \subfigure[Minimax, nominal]{
%% %% \includegraphics[width=6.5cm]{mm_example1_dose_s0}
%% %% }
%% %% \subfigure[Minimax, perturbed]{
%% %% \includegraphics[width=6.5cm]{mm_example1_dose_s4}
%% %% }
%% %% \subfigure[Expected-value, nominal]{
%% %% \includegraphics[width=6.5cm]{expval_example1_dose_s0}
%% %% }
%% %% \subfigure[Expected-value, perturbed]{
%% %% \includegraphics[width=6.5cm]{expval_example1_dose_s4}
%% %% }
%% \subfigure[Scenario-based margins]{
%% \includegraphics[width=6.5cm]{gm_example1_dose_s0}
%% }
%% %\subfigure[Scenario-based margins, perturbed dose]{
%% %\includegraphics[width=6.5cm]{gm_example1_dose_s4}
%% %}
%% \subfigure[Scenario-based margins{
%% \includegraphics[width=6.5cm]{gm_example1_beamdose_s0}
%% }
%% \caption{Dose distributions for Example 2. (a): Nominal total dose. (b): Total dose in the scenario where the beams are shifted by $1$ cm in both the positive x-direction and the positive y-direction. (c): Nominal beam dose for the beam along the y-axis.}
%% \label{fig-2d-1-dose}
%% \end{figure}

\subsection{Example 3}
The purpose of this example is to investigate how scenario-based margins handle a situation when target coverage in some scenarios is incompatible with OAR sparing. The incompatibility stems from that the PTV overlaps with the PRV. The static dose cloud approximation is close to exact: it would have been exact had it not been for shift scenarios in pairwise combinations of the axis direction that do not line up exactly with the dose grid. The results for geometric margins were, nevertheless, highly similar to the results for scenario-based margins ($98 \p$ of the differences in voxel dose were less than $3 \p$ of the prescription) and are therefore not shown.

\begin{figure}[hbtp]
\centering
%\subfigure[Margin]{
%\includegraphics[width=6.5cm]{margin_example2_dose_s0}
%}
\subfigure[Scenario-based margins]{
 \label{fig-2d-2-dose-a}
\includegraphics[width=4.3cm]{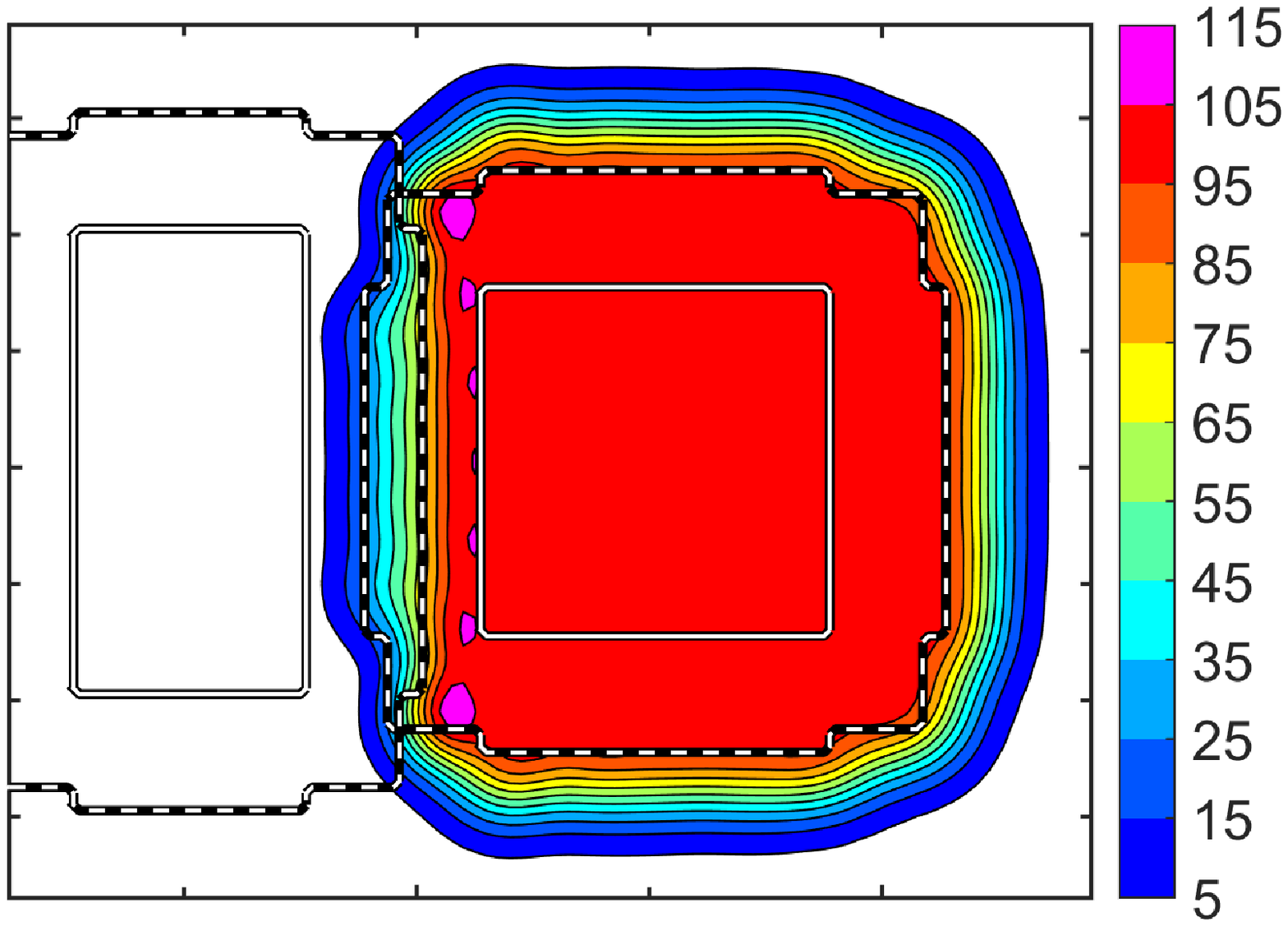}
}
\subfigure[Worst case optimization]{
 \label{fig-2d-2-dose-b}
\includegraphics[width=4.3cm]{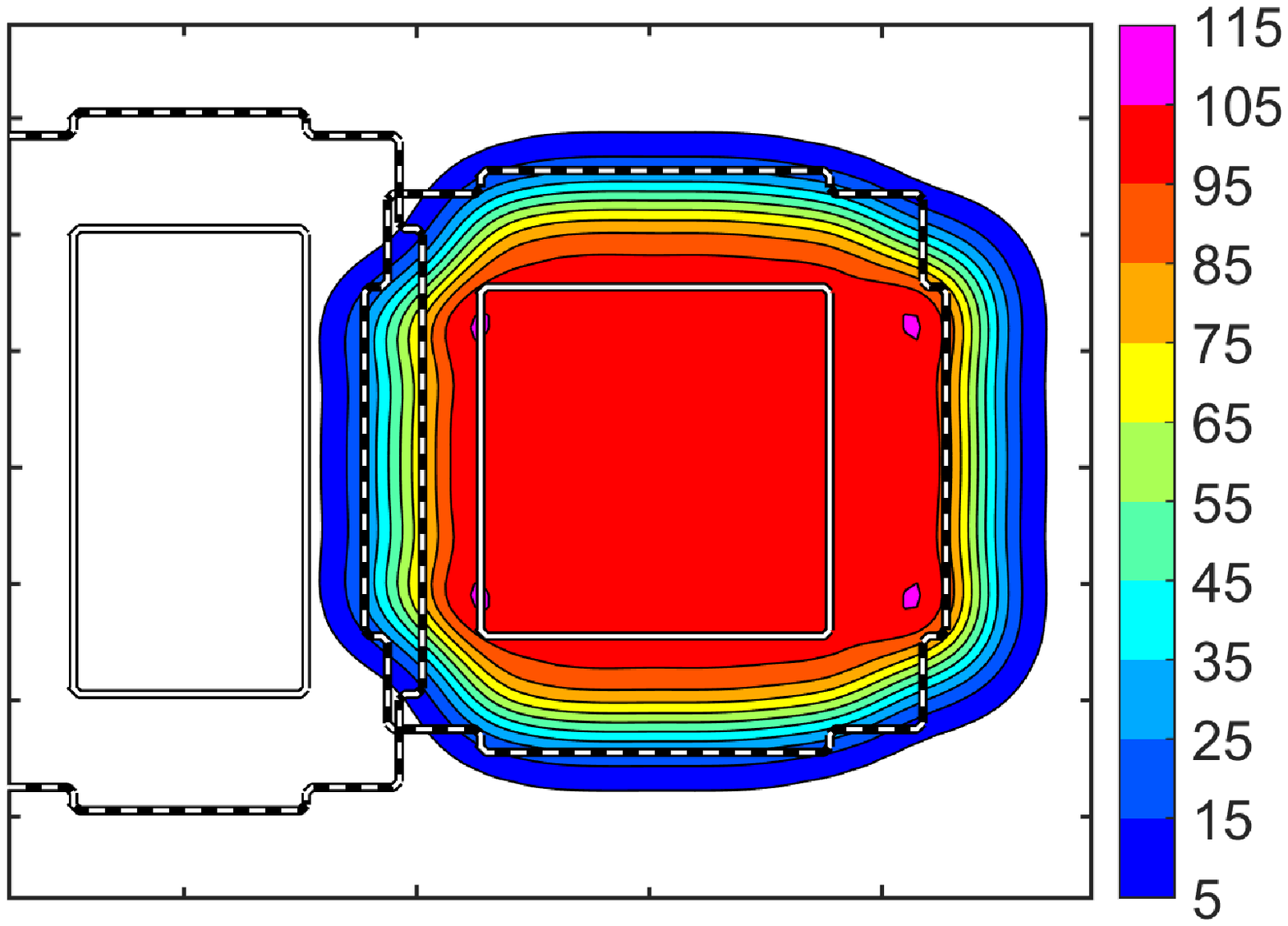}
}
\subfigure[Expected value optimization]{
 \label{fig-2d-2-dose-c}
\includegraphics[width=4.3cm]{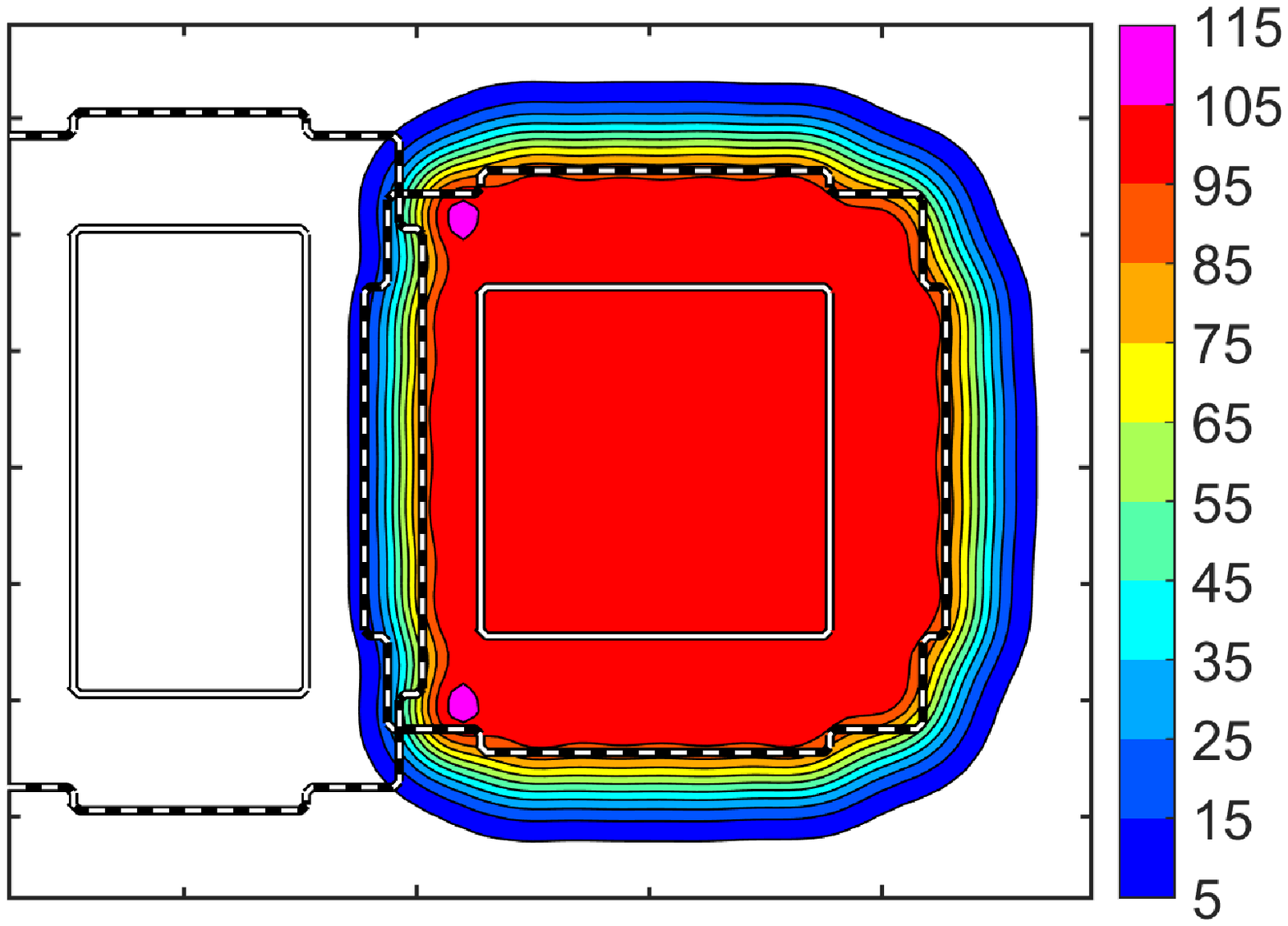}
}
\caption{Nominal doses for Example 3. The CTV and OAR are indicated by solid lines and the PTV and PRV are indicated by dashed lines.}
\label{fig-2d-2-dose}
\end{figure}

\begin{figure}[hbtp]
\centering
\includegraphics[width=6.5cm]{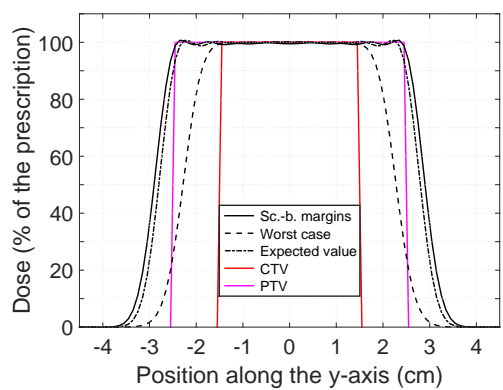}
\caption{Line dose profiles for Example 3 in the nominal scenario for a trace along the y-axis through the isocenter. The height of the CTV and PTV indicates the prescription dose.}
\label{fig-2d-2-linedose}
\end{figure}

Figure~\ref{fig-2d-2-dose} shows that margins and expected value optimization extended the high-dose region from the CTV to the edges of the PTV except for in the leftward direction that is in conflict with the OAR. The CTV coverage was therefore maintained under all shifts without rightward component. Coverage was lost almost completely for $20\p$ of the CTV under rightward shifts, see Figure~\ref{fig-2d-2-dvh}. A minor difference for expected value optimization compared to margins was that the doses in the CTV-to-PTV margin were slightly lower and the dose fall-off was more gradual. Worst case optimization only extended the high-dose region to the right of the CTV. The CTV coverage was therefore poor not only under the rightward shifts that are in conflict with OAR sparing, but also under the relatively ``easier'' up- and downward shifts. The line doses in Figure~\ref{fig-2d-2-linedose} demonstrate how the high-dose region for worst case optimization was retracted in the up- and downward directions compared to the other methods. 

\begin{figure}[hbtp]
\centering
%\subfigure[Margin]{
%\includegraphics[width=4.3cm]{margin_example2_dvh}
%}
\subfigure[Scenario-based margins]{
  \label{fig-2d-2-dvh-a}
\includegraphics[width=4.3cm]{gm_example2_dvh}
}
\subfigure[Worst case optimization]{
  \label{fig-2d-2-dvh-b}
\includegraphics[width=4.3cm]{mm_example2_dvh}
}
\subfigure[Expected value optimization]{
  \label{fig-2d-2-dvh-c}
\includegraphics[width=4.3cm]{expval_example2_dvh}
}
\caption{DVHs for Example 3 in all scenarios.}
\label{fig-2d-2-dvh}
\end{figure}

\subsection{Example 4}
The difference between this example and Example 3 is that the PRV overlaps with the CTV and the weight on the OAR objective is five times smaller than previously. The results for geometric margins are not depicted because of close similarity with the results for scenario-based margins ($98\p$ of the differences in voxel dose were less than $3\p$ of the prescription). 

\begin{figure}[hbtp]
\centering
%\subfigure[Margin]{
%\includegraphics[width=6.5cm]{margin_example2_dose_s0}
%}
\subfigure[Scenario-based margins]{
  \label{fig-2d-3-dose-a}
\includegraphics[width=4.3cm]{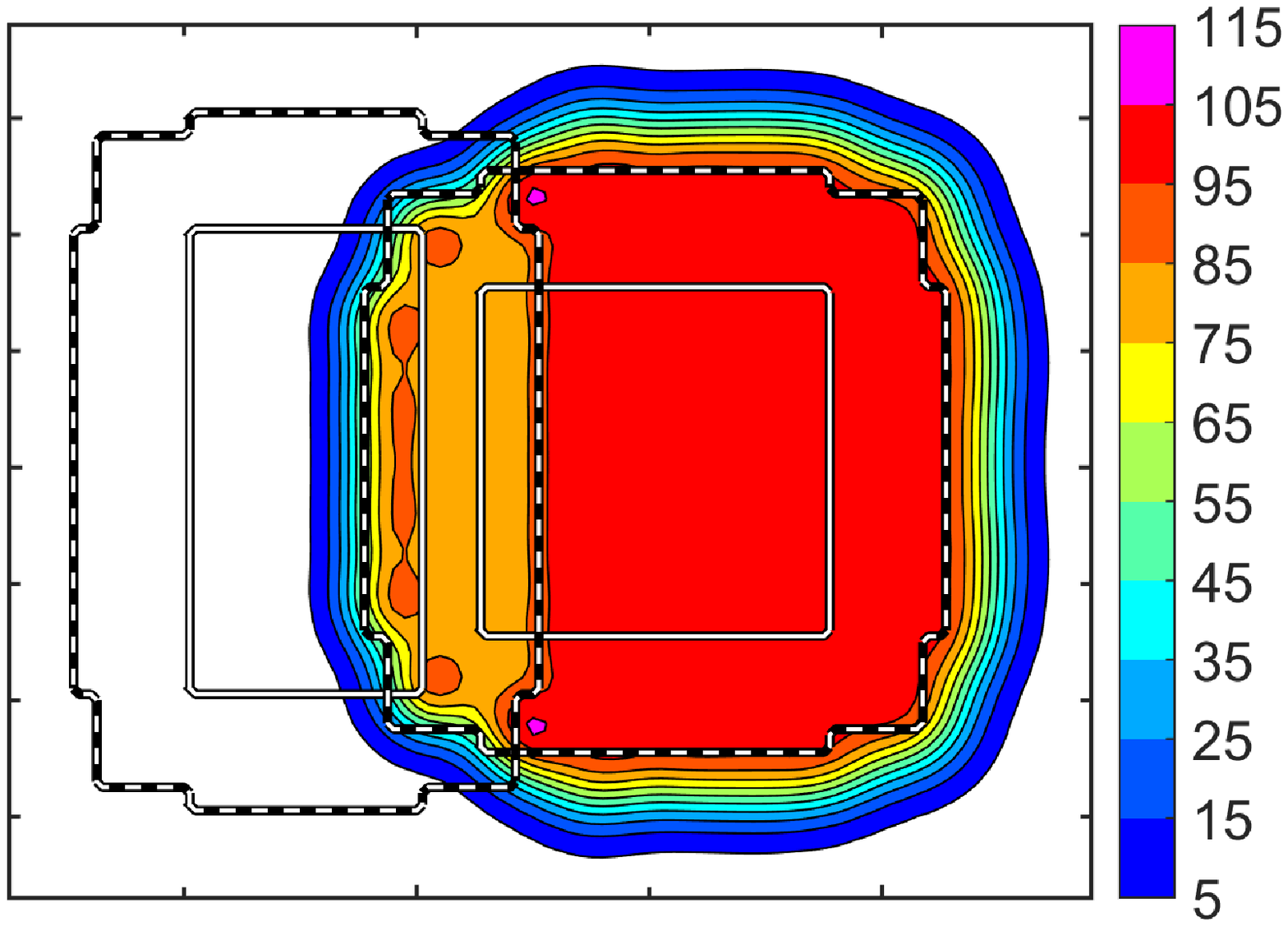}
}
\subfigure[Worst case optimization]{
  \label{fig-2d-3-dose-b}
\includegraphics[width=4.3cm]{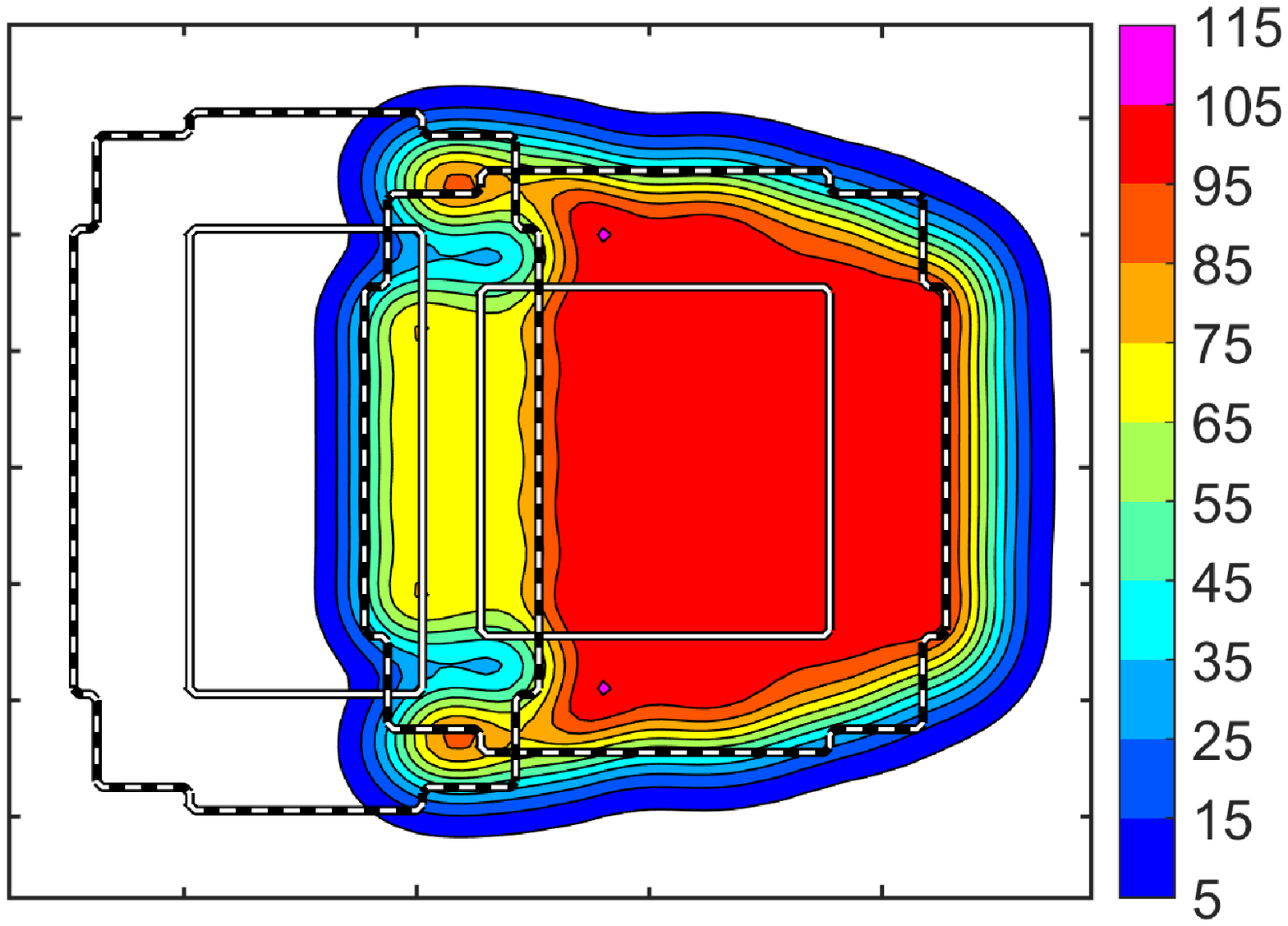}
}
\subfigure[Expected value optimization]{
  \label{fig-2d-3-dose-c}
\includegraphics[width=4.3cm]{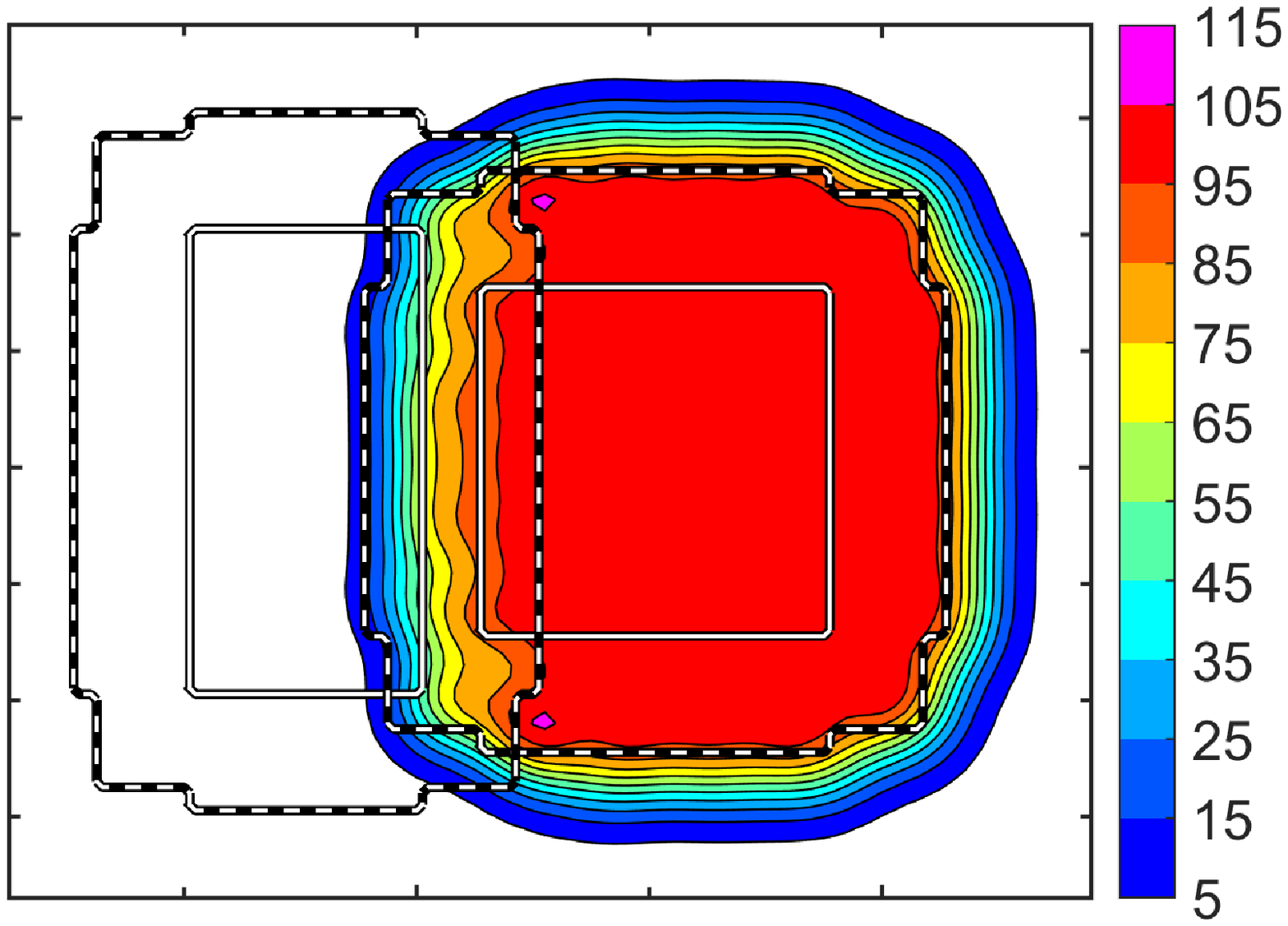}
}
\caption{Nominal dose distributions for Example 4. The CTV and OAR are indicated by solid lines and the PTV and PRV are indicated by dashed lines..}
\label{fig-2d-3-dose}
\end{figure}

\begin{figure}[hbtp]
\centering
\includegraphics[width=6.5cm]{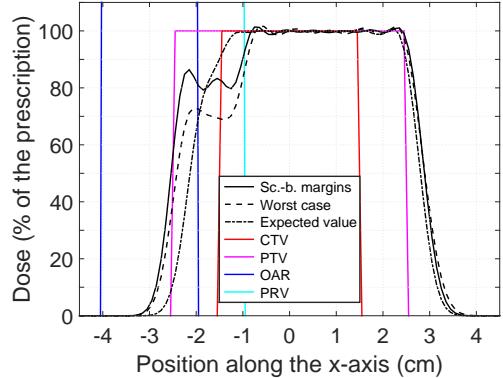}
\caption{Line dose profiles for Example 4 in the nominal scenario for a trace along the x-axis through the isocenter. The height of the CTV and PTV indicates the prescription dose. The OAR and PRV boundaries are indicated by vertical lines.}
\label{fig-2d-3-linedose}
\end{figure}

The lesser priority on OAR sparing compared to Example 3 made margins accept a dose of about $80\p$ in the overlap region between the PTV and PRV, see Figure~\ref{fig-2d-3-dose}. The dose compromise in this region was controlled in the sense that the dose was uniform, except for peaks in the dose at the left edge of the PTV. Worst case optimization gave a uniform dose at the prescription level to the CTV in the left shift scenario, and underdosed the CTV to about $70\p$ in the region where it overlaps with the PRV. The CTV coverage was poor also under all shifts with an up- or downward component because there was little extension of the high-dose region outside the CTV in these directions, see Figure~\ref{fig-2d-3-dvh}. Expected value optimization covered almost the entire CTV with a dose of one in the nominal scenario, thereby accepting doses above $90\p$ to more than $20\p$ of the OAR if a leftward shift occurs. The CTV dose was highly non-robust to shifts in the right direction because the dose falls gradually towards zero left of the CTV edge, see Figure~\ref{fig-2d-3-linedose}. The CTV coverage under a shift with leftward component was comparable to margins. 

\begin{figure}[hbtp]
\centering
%\subfigure[Margin]{
%\includegraphics[width=4.3cm]{margin_example2_dvh}
%}
\subfigure[Scenario-based margins]{
  \label{fig-2d-3-dvh-a}
\includegraphics[width=4.3cm]{gm_example3_dvh}
}
\subfigure[Worst case optimization]{
  \label{fig-2d-3-dvh-b}
\includegraphics[width=4.3cm]{mm_example3_dvh}
}
\subfigure[Expected value optimization]{
  \label{fig-2d-3-dvh-c}
\includegraphics[width=4.3cm]{expval_example3_dvh}
}
\caption{DVHs for Example 4 in all scenarios.}
\label{fig-2d-3-dvh}
\end{figure}

\section{Discussion}
%In this section, we first review the benefits of a scenario-based optimization in view of Example 2. We then discuss the benefits of scenario-based margins compared to previous scenario-based methods in view of Examples 1, 3 and 4. Finally, we comment on the disadvantages of scenario-based margins and outline directions for future research.

\subsection{Advantages over geometric margins}
There is clear evidence that scenario-based optimization can provide better target coverage robustness than margins and also reduce the dose to healthy tissues if the static dose cloud approximation is inaccurate~\cite{fredriksson11,fredriksson12,chen12,liu12}. Scenario-based optimization is therefore particularly advantageous for ion therapy, treatments of heterogeneous densities, and treatments with few fields. Our results for Example 2 reinforce this conclusion: geometric margins created nonuniform beam doses that were sensitive to perturbations and placed more dose inside the CTV-to-PTV margin than necessary, while all scenario-based methods created highly robust plans. Furthermore, a potential advantage of scenario-based methods, even if scenario doses are calculated using the static dose cloud approximation, is that robustness can be implemented for nonuniform CTV prescriptions, as in dose-painting guided by functional imaging~\cite{bentzen05}.

\subsection{Advantages over other scenario-based models}
\subsubsection{Dose gradients}
This study, alongside previous data~\cite{fredriksson12}, shows that expected value optimization with respect to systematic setup uncertainty can lead to doses that fall gradually from the prescription level to a dose of zero over a prolonged distance outside the CTV. Such a gradual fall-off was observed consistently in Examples 1, 3, and 4. Margins and worst case optimization instead created dose shoulders at the PTV boundary that contributed to a steep fall-off. The gentle fall-off created by expected value optimization is a point of concern in view of the fact that the possibility to deliver dose-escalated treatments without compromising the dose to surrounding organs has been a key driver in the adoption of intensity-modulated over conformal radiation therapy~\cite{sheets12}: van Herk et al.~\cite{vanherk02} showed that steep dose gradients are necessary for such dose escalation. Proponents of proton therapy also often point to the steep fall-off on the lateral and distal edge of the target as evidence of its superiority over photon treatments~\cite{suit08}. 

\subsection{Tradeoffs between conflicting criteria}
In a previous study~\cite{fredriksson14}, we showed that worst case optimization can be too conservative if target coverage to some extent is incompatible with sparing of an OAR. The optimization can then neglect coverage also in scenarios that are not in direct conflict with the OAR. This effect was also apparent in Examples 3 and 4, where worst case optimization disregarded CTV coverage with respect to shifts with an up- or downward component. Margins and expected value optimization did not exhibit this shortcoming. These methods instead created robust CTV coverage for all shifts except those that required an extension of the high-dose region into the PRV. For the worst case optimization method to work as intended, the scenarios of non-resolvable conflict could be retracted.

Margins and expected value optimization exhibited markedly different behaviors with respect to the dose compromise in the overlap region between the PTV and PRV in Example 4. Margins created a controlled underdosage of the CTV while expected value optimization sacrificed CTV coverage under a rightward shift and OAR sparing under a leftward shift. This is an effect of that the nominal scenario has a larger probability of occurring than the shift scenarios. Furthermore, the results for expected value optimization depend on the scenario discretization: The probability for the extreme shift scenarios become smaller if intermediate shifts are also incorporated in the uncertainty set $\S$ due to normalization of the discretized probabilities.  

It is a clinical decision whether the characteristics of margins or expected value optimization are preferable. The current ICRU guidelines advocate that when the PTV and OAR or PRV overlap, priority rules should be used, or the regions should be subdivided into regions with different absorbed-dose constraints~\cite[Section 4.5]{icru83}. This effect is achieved by margin-based planning with importance weights for the PTV and the OAR or PRV used in the optimization. Several clinical protocols, however, prescribe a smaller margin in the interface against some radiation sensitive structures, for example a retracted CTV-to-PTV margin for prostate in the posterior direction to promote rectal sparing~\cite{zelefsky99, dearnaley12}. It is our view that if a retraction is to be made in a scenario-based model, it should be incorporated in the selection of scenarios rather than in their probabilities, in order to maintain steep dose gradients. Optimization of the definition of the uncertainty set $\S$ subject to constraints that a set of clinical goals is satisfied for all scenarios in $\S$ is described in Fredriksson et al.~\cite{fredriksson15}.

\subsection{Potential disadvantages}
\subsubsection{Nonconvexity}
The scenario-based margins lead to a nonconvex optimization model if beam-specific effects are taken into account, as shown in Appendix~\ref{app-proof}. Nonconvex optimization problem can have multiple local minima, meaning that optimization algorithms cannot guarantee convergence to the global optimum. While a nonconvex formulation allows the existence of multiple local optima, however, it does not assure that such optima exist. The local optima may also be of negligible clinical difference even if they do exist.

\subsubsection{Limitation to setup errors in homogeneous media}
The beam-specific form of the scenario-based margins does not take the mass density of the treatment volume into account in the calculation of the beam-, scenario- and voxel-specific weights. The model is therefore appropriate for treatments with few fields, as demonstrated by Example 2, but although the correct scenario doses are used, the weights may be inappropriate for irradiation of heterogeneous media. The scenario-based margins as presented here are also not designed for handling of ion beam range errors, and cannot utilize multiple patient images for handling of inter- and intra-fraction organ motion. Extensions to these types of uncertainty will be presented in a separate publication.

\section{Conclusions}
We have introduced a scenario-based generalization of treatment plan optimization with margins to handle setup uncertainties. When the static dose cloud approximation is exact, the generalization is equivalent to margins, but when the approximation is inaccurate, the generalization exhibits the benefits of scenario-based optimization. Comparative planning on phantom geometries demonstrated that the scenario-based margins can provide robustness where geometric margins fail and at the same time avoid some disadvantages of previous methods for robust planning, such as the unnecessarily gentle dose fall-off of expected value optimization and the neglection of ``easy'' scenarios of worst case optimization.
	
\appendix
\section{Proof of nonconvexity}\label{app-proof}
The following example shows that a beam-weighted criterion $f$ defined according to~\eqref{beam-weighting} can be nonconvex even when $\phi$ is convex. Consider a 2D geometry where a single voxel is irradiated by a first beam along the x-axis and a second beam along the y-axis. Let $\phi(d) = \abs{d}^{3/2}$ and let $\S$ be an uncertainty set composed of the nominal scenario and a 1-voxel shift along the y-axis. The beam- and scenario-specific weights then take on the values $p_{s}^{(1)} = 1$ and $p_{s}^{(2)} = 1/2$ for all $s$ in $\S$. Consider two plans A and B with beam doses according to
\[
\quad \dA^{(b)}(s) = \left\{ \begin{array}{ll} 2 & \textrm{if $b = 1$} \\ 0 & \textrm{otherwise} \end{array} \right. \quad \textrm{and} \quad \dB^{(b)}(s) = \left\{ \begin{array}{ll} 4 & \textrm{if $b = 2$} \\ 0 & \textrm{otherwise} \end{array} \right., \quad \forall s \in \S. 
%\quad d_X^{(b)}(s) = \left\{ \begin{array}{ll} 2 & \textrm{if $X$ = A and $b = 1$} \\ 4 & \textrm{if $X$ = B and $b = 2$} \\ 0 & \textrm{otherwise} \end{array} \right. \quad \forall s \in \S.
\]
%\[
%p_{s}^{(b)} = \left\{ \begin{array}{ll} 1 & \textrm{if $b = 1$} \\ 1/2 & \textrm{otherwise} \end{array} \right.. 
%\]
%for all $s$ in $\S$.
Convexity of $f$ requires that the function value of the average dose is less than or equal to the average function value. The inequality
\[
f((\dA+\dB)/2) = 4\sqrt{3} > (f(\dA) + f(\dB))/2 = 2(\sqrt{2}+2),
\]
therefore, proves the nonconvexity of $f$.

%$p_{s_1}^{(1)} = p_{s_2}^{(1)} = 1$ and $p_{s_1}^{(2)} = p_{s_2}^{(2)} = 1/2$. Consider two plans A and B such that $d_{A}^{(1)}(s_1) = d_{A}^{(1)}(s_2) = 2$, $d_{B}^{(2)}(s_1) = d_{B}^{(2)}(s_2) = 4$, and $d_{A}^{(2)}(s_1) = d_{A}^{(2)}(s_2) = d_{B}^{(1)}(s_1) = d_{B}^{(1)}(s_2) = 0$. 

%\[
%\sum_{s \in \S} \sum_{b \in \B} \frac{d^{(b)}(s)}{d(s)} p_{s}^{(b)} \phi( d(s) ),
%\]
%
%\[
%f(A) =  \underbrace{\frac{2}{2} \times 1 \times 2^{3/2}}_{b1, s1} +
%        \underbrace{\frac{0}{2} \times 1/2 \times 2^{3/2}}_{b2, s1} +
%        \underbrace{\frac{2}{2} \times 1 \times 2^{3/2}}_{b1, s2} +
%        \underbrace{\frac{0}{2} \times 1/2 \times 2^{3/2}}_{b2, s2}
%= 4 \sqrt{2}
%\]
%
%\[
%f(B) =  \underbrace{\frac{0}{4} \times 1 \times 4^{3/2}}_{b1, s1} +
%        \underbrace{\frac{4}{4} \times 1/2 \times 4^{3/2}}_{b2, s1} +
%        \underbrace{\frac{0}{4} \times 1 \times 4^{3/2}}_{b1, s2} +
%        \underbrace{\frac{4}{4} \times 1/2 \times 4^{3/2}}_{b2, s2}
%= 8
%\]
%
%\[
%(f(A)+f(B))/2 = 2 (\sqrt{2} + 2 ) \approx 6.828
%\]
%
%\[
%f((A+B)/2) =  \underbrace{\frac{1}{3} \times 1 \times 3^{3/2}}_{b1, s1} +
%        \underbrace{\frac{2}{3} \times 1/2 \times 3^{3/2}}_{b2, s1} +
%        \underbrace{\frac{1}{3} \times 1 \times 3^{3/2}}_{b1, s2} +
%        \underbrace{\frac{2}{3} \times 1/2 \times 3^{3/2}}_{b2, s2}
%= 4 \sqrt{3} \approx 6.928
%\]
%

%\bibliography{bibliography}
%\bibliographystyle{bibstyle}

\end{document}